\documentstyle[11pt]{article}
\begin{document}

\title{Cosmological quintessence accretion onto primordial black holes :
conditions for their growth to the supermassive scale }

\author{P. S. Cust\'odio and J.E.Horvath,\\
\it Instituto de Astronomia, Geof\'\i sica e Ci\^encias Atmosf\'ericas\\
\it IAG/USP, Rua do Mat\~ao, 1226, 05508-900 S\~ao Paulo SP,
Brazil\\ Email: foton@astro.iag.usp.br}
\maketitle

\vskip5mm

\abstract{In this work we revisit the growth of small primordial
black holes (PBHs) immersed in a quintessential field and/or
radiation to the supermassive black hole (SMBHs) scale. We show
the difficulties of scenarios in which such huge growth is
possible. For that purpose we evaluated analytical solutions of
the differential equations (describing mass evolution) and point
out the strong fine tuning for that conclusions. The timescale for
growth in a model with a constant quintessence flux is calculated
and we show that it is much bigger than the Hubble time.The
fractional gain of the mass is further evaluated in other forms,
including quintessence and/or radiation. We calculate the
cosmological density $\Omega$ due to quintessence necessary to
grow BHs to the supermassive range and show it to be much bigger
than one. We also describe the set of complete equations analyzing
the evolution of the BH+quintessence universe, showing some
interesting effects such the quenching of the BH mass growth due
to the evolution of the background energy. Additional constraints
obtained by using the Holographic Bound are also described. The
general equilibrium conditions for evaporating/accreting black
holes evolving in a quintessence/radiation universe are discussed
in the Appendix.}

\section{Introduction}

Primordial black holes may be important in a number of situations.
If PBHs formed in the early universe from primordial perturbations
or some other mechanism, it is still not excluded that a
population may survive. Some of them may even become the
supermassive black holes that energize active galactic nuclei
driving the powerful source of quasars, and may be a powerful
sources of gravitational waves as well (see the review of M.Rees
\cite{MR}).

The origin of supermassive black holes (SMBHs) remains a mystery,
and since there are robust observational evidence for their
existence with huge masses
(${10}^{6}{M_{\odot}}-{10}^{8}M_{\odot}$), a big problem arises.
How these objects can be formed in the universe? Were their
initial masses very large from the beginning?  There are several
proposals for understanding these questions. One can suppose that
the big black holes lurked in the AGNs can be formed from a
primordial end point of the first generation of stars. A galactic
nucleus could be populated by several massive bodies, and these
objects fused into a only and single massive collapsed body,
forming a central black hole. These models are discussed, for
example, in Refs.\cite{Ty} and references therein.

A second class of models (studied here) assume a huge growth of
PBHs as a working hypothesis. Quite independently of the formation
scenarios of small black hole seeds, the important question is
whether they can grow to the supermassive range by means of
accretion.

The analysis of the Hawking evaporation \cite{SH} of PBHs has
leaded to conclude that PBHs initially larger than
$M_{haw}\sim{10}^{15}g$ could have chance to survive and
contribute to a small fraction of the dark matter. A subset of
PBHs may have grown to the supermassive range residing within the
observed AGNs. In Ref. \cite{CuHo} we evaluated in detail the
evolution and survival of PBHs taking into account evaporation and
absorption from a radiation-dominated environment. We proved that
all those PBHs within in the radiation-dominated era remained
colder than the environment, and had a very small mass growth.

The simplest possibility for reaching the supermassive range is
that PBHs formed in the early universe accreted mass from an
energetic radiation environment (the Hawking evaporation was
important only when their proper temperature became higher than
the background temperature). If PBHs were massive enough, they
could be colder than the environment, and accretion will drive
their evolution. The accretion term may be constructed by
multiplying the cross section for classical absorption of
relativistic particles by the flux coming through the horizon.
Taking into account the gravitational focusing of the incident
radiation, this quantity is of order

\begin{equation}
\biggl({dM\over{dt}}\biggr)_{abs}
={27\pi\over{4}}{R_{g}}^{2}c\varrho_{rad}(T) ;
\end{equation}

where the factor $27\pi/4$ describes the relativistic beaming of
particles, see Zel'dovich and Novikov for details \cite{ZN}. Since
the gravitational radius is $R_{g}=2GM/{c}^{2}$, this accretion is
proportional to the second power of the black hole mass $M$. The
formula above assumes that (in a first approximation) classical
accretion does not affect the metric that describes the black hole
itself. The issue is whether one can construct quintessence models
which not only give the correct dynamics for the expansion, but
also make PBHs grow to the supermassive range. This possibility is
the subject of the following Sections.

\bigskip
\section{Quintessence models and Supermassive Black Hole growth}
\bigskip

There is now very good evidence that a dark energy is present in
our universe and represents $\sim{0.7}$ of the critical density.
Today, the big challenge is understand the physical nature of this
contribution. A crucial ingredient is that, contrary to the dark
matter, dark energy must possess a negative pressure as required
for an accelerating universe. Then, $\varrho_{eff}=(\varrho+3P) <
0$, implies $w \, < \, -1/3$, with $w \, = \, P/\varrho$ the
velocity of sound.

At first sight, the most natural candidate is the cosmological
constant. The cosmological constant can be viewed as a fluid
exactly possessing $w \, = \, -1$. Since
 $\dot{\varrho}+3H(1+w)\varrho=0$, implies that
 $\varrho \, = \, constant$. However,
as it is well-known, explaining the dark energy with a
cosmological constant runs into problems. The first, is that we
need to generate a tiny (``unnatural") value at present epoch, to
explain the magnitude of the acceleration of the universe.  The
second problem is the so-called "coincidence" problem. At high
redshifts, say just after inflation ($z\sim{10}^{28}$), the
radiation density was $\varrho_{rad}\sim{10}^{61}GeV^{4}$, while
$\varrho_{\Lambda}\sim{10}^{-47}GeV^{4}$. This means that in order
to have the correct amount of dark energy today, we need to
fine-tune the initial conditions such that the ratio above was
${10}^{110}$ with a very high precision. This also seems very
unnatural. These considerations leaded to consider that the dark
energy is time-dependent. As is known from inflation models, a
simple way of obtaining these features is to consider a minimally
coupled scalar field $\varphi$ (termed quintessence). The latter
is {\it not} necessarily the inflaton field, because we are at a
much smaller energy scale, although the dynamical equations can be
similar. Since the scalar field evolves with time, the equation of
state itself varies with redshift, and it constitutes a very
important observational challenge to measure this dependence
\cite{BM}.

It is beyond the scope of our work to analyze the whole features
of the quintessential models, but instead we elaborate on the
suggestion of Bean and Magueijo \cite{BeMa}, who suggested an
important growth of seed PBHs when a quintessence scalar field
$\varphi$ is absorbed, dominating the accretion. The key new
ingredient is the role played by the kinetic term
$({{{\dot{\varphi}}^{2}\over{2}}})$ in the flux onto the PBHs,
which is absent in the case of accretion of pure radiation. They
further showed that the flux at the horizon is given by
$F_{\varphi}(t)={{{\dot{\varphi}}^{2}\over{2}}}(t)$ (in natural
units), and the potential term $V(\varphi(t))$ drives the behavior
of the kinetic term through the dynamical equations describing
conservation and the cosmic expansion.

Type Ia supernovae data suggests that at $z < 1$ the dark energy
is dominant \cite{Super}. It is sure that for $z \gg 1$, the dark
energy was not dominant, because eventually matter becomes
important at these redshifts and earlier, for $z \, \sim \,
10^{4}$, radiation becomes more important. Then, we can conclude
that, although we can not yet measure the dark energy contribution
at $z \geq 1$, that contribution is subdominant but may be not
negligible for the black hole growth problem from epochs earlier
than present. We can view this in other form. The complete
dynamics of quintessence plus radiation (and some black holes
mixed as a secondary dark contribution) shows attractor solutions.
The simplest models with quintessence show that it is very hard to
have at, say, $z \sim {10}$ and $\Omega_{DE}<<1$, while today
$\Omega_{DE}(z\sim{1})\sim{0.7}$. There are not ``natural"
solutions containing these features.

We shall describe a system considering that the energy and
pressure of the quintessential field dominate the black hole
contribution, and one simplified model arises from there.

A general model containing black holes + quintessence of the type
studied here is described by

\begin{equation}
S=\int{d^{4}x\sqrt{-g}}({\kappa}R+L_{\varphi}+L_{matter}) \,\,\, ;
\end{equation}

where the quintessence scalar field has pressure and
energy-density given by
$P(t)={\dot{\varphi}^{2}\over{2}}-V(\varphi)$ and
$\varrho(t)={\dot{\varphi}^{2}\over{2}}+V(\varphi)$ respectively.

We assume that our spacetime is asymptotically flat, consisting of
this quintessential field and black holes (without charge or
angular momentum) absorbing the energy of this field and neglect
all other contributions, here $\kappa=1/16\pi{G}$  is the
gravitational constant. The total $T_{\mu\nu}$ does not need to be
diagonal, even if the metric is diagonal in the beginning. The
non-diagonal terms describe energy-flux at black hole horizons.

The dynamics of the system is given by

\begin{equation}
\biggl({dM\over{dt}}\biggr)_{growth}={CM^{2}\over{2}}{\dot{\varphi}^{2}}
\,\, ;
\end{equation}

\begin{equation}
\ddot{\varphi}+3H\dot{\varphi}+V^{\prime}= 0 \,\, ;
\end{equation}

\begin{equation}
H^{2}={\biggl({\dot{a}\over{a}}\biggr)}^{2}={8\pi\over{3{M_{pl}}^{2}}}
(V+\varrho_{pbh}) -{K\over{a^{2}}} \,\,\,;
\end{equation}

\begin{equation}
\biggl({\ddot{a}\over{a}}\biggr)=-{4\pi\over{3M_{pl}^{2}}}(\varrho+3P)
\,\,\, ;
\end{equation}

where $a(t)$ denotes the scale factor of expansion/contraction of
the universe and $K \, = \, -1,0,+1$ describes the cosmic
curvature. We shall evaluate physical effects in the flat model $K
\, = \, 0$ for simplicity, and assume that all the black holes
started with the same mass, i.e. a single-mass scale initial
function. We use natural units in which $C=27\pi/{M_{pl}}^{4}$.
Note that eq.(4) is the conservation law for the quintessential
field in the form $\dot{\varrho}+3H(\varrho+P) = 0$.

Since we have assumed from the scratch that PBHs do not dominate
the dynamics of the universe $\varrho_{pbh}<<V(\varphi)$,
analytical solutions for $\varphi(t)$ and its derivatives can be
obtained without additional complications. It is obvious that,
when this requirement breaks down, $\varphi(t)$ becomes a very
complex function of $t$ and $M(t)$ and a numerical analysis is
necessary.

Bean and Magueijo \cite{BeMa} have recently analyzed only one
particular model of quintessence in which
$V(\varphi)=\lambda{exp[\lambda{\varphi}]}$, implying
$\dot{\varphi}^{2}\propto{t}^{-2}$ within the slow-roll
approximation. Within this model, in which the quintessence flux
around the SMBH {\it decreases} with time, they claim that the
black hole mass grows with growing time. However, the decreasing
quintessence flux around these objects ($\propto{t}^{-2}$)implies
that the mass gain term decreases accordingly
$(\dot{M}\propto{M^{2}{t}^{-2}})$. In fact, we have proved that in
the radiation-dominated era the growth of black holes is quenched
when the background flux decreases with time as ${t}^{-2}$ or
faster (this argument is generalized in the present work for
quintessence accretion, see also Ref.\cite{CuHo1}). It is
important to gain some feeling about the behavior of simple
solutions first, and we start with one of the simplest
possibilities, namely that the flux of quintessence onto the black
holes remains constant.

\subsection{Constant quintessence-flux through the horizon event}

Since we are interested in evaluating the effect of the
quintessential field on the black hole growth, we may start
solving the particular case in which the kinetic term (and hence
the flux)  $F(t) \propto {\dot{\varphi}}^{2}$ is constant. If we
set ${\dot{\varphi}}= constant \equiv \, B$, eq.(4) yields

\begin{equation}
{{\sqrt 48 \pi\alpha(B)}\over{M_{pl}}} = - {dV\over{V^{1/2}}}
\,\,\, ;
\end{equation}

with $\alpha(B)={\sqrt{48\pi{B}}\over{M_{pl}}}$ and a slow-roll
dynamics of the quintessence field holds. Solving for $V$ yields

\begin{equation}
V(\varphi)=V_{0}\sqrt{1+{\alpha(B)}(\varphi-\varphi_{0})} \,\,\, ;
\end{equation}

with $\alpha(B)=-({12 \pi B \over{M_{pl}^{2}}})^{1/2}$.

The scalar field itself evolves as $\varphi(t)=\varphi_{0}
+\sqrt{2B}(t-t_{0})$. Thus, the explicit solution for the
evolution of the mass of the black hole is

\begin{equation}
M(t)={M_{i}\over{\biggl[{1-{54\pi{M_{i}}B\over{M_{pl}^{4}}}(t-t_{i})}}\biggr]}
\,\,\, .
\end{equation}

For $t \gg t_{i}$, eq.(9) becomes

\begin{equation}
M(t)={M_{i}\times{\biggl[{1-{2.8\times{10}^{-22}}{h_{0}}^{-1}
\biggl({B\over{\varrho_{c}}}\biggr)\mu\tau }\biggr]}^{-1}} \,\,\,
;
\end{equation}

where $\mu=(M_{i}/M_{\odot})$, $\tau=(t/t_{0})$ and
$\varrho_{c}(t_{0})\sim{10}^{-29}g{cm}^{-3}$.

It is clear that the asymptotic value of the mass is much larger
than the initial mass $M(t) \, \gg \, M_{i}$ if the denominator
above satisfies

\begin{equation}
0\, < \, \biggl[{1-{1-{2.8\times{10}^{-22}}{h_{0}}^{-1}
\biggl({B\over{\varrho_{c}}}\biggr)\mu\tau }}\biggr] \, < \delta
\,\,\, .
\end{equation}

with $\delta \ll 1$. The PBH mass actually diverges at
$t_{a}(M_{i})={M_{pl}^{4}\over{54\pi{B}M_{i}}}$, numerically given
by

\begin{equation}
t_{a}(M_{i},B) \sim  \, 3.4 \times{10}^{21} {\biggl[{
\varrho_{c}(t_{0})\over {B}}\biggr]} \times {\biggl[
{M_{\odot}\over {M}} \biggr]} \, t_{0} \,\,\, ;
\end{equation}

with $t_{0}\sim{10}^{17}{h_{0}}^{-1}s$ and
$\varrho_{c}(t_{0})\sim{10}^{-29}g{cm}^{-3}$. This of course does
{\it not} mean that a black hole will achieve an arbitrary large
mass, but rather that before this happens, our approximations will
break down and the condition ${\dot{\varphi}}= constant$ will not
be satisfied. An analogous exercise in which the quintessence flux
decreases as $F(t) \propto{t}^{-1}$ still produces a (slow but
reasonable) growth over the cosmic time. However, for
$F(t)\propto{t}^{-n}$, with $n > 1$ the growth is quenched because
the environment does not provide enough fuel as time goes by (see
Section 3).

The next natural question is: how do the PBHs avoid the divergence
of their masses in a "natural" way? For a generic quintessential
field, but keeping the general slow-roll approximation for the
dynamics of $\varphi$, we use the eqs.(3), (4) and (5) to obtain
the mass of the black hole as a solution of a more complicated
equation

\begin{equation}
\biggl({dM\over{dt}}\biggr)={C{M_{pl}}^{2}\over{48\pi}}
{M^{2}{(V^{\prime}(\varphi(t))+{\ddot{\varphi}})}^{2}\over{V(\varphi(t))+\varrho_{pbh}(t)}}
\,\,\, .
\end{equation}

The non-trivial effects of the presence of other black holes are
now explicitly seen. In the limit $\varrho_{pbh} > V(\varphi)$ it
is expected that the flux at the horizon decreases and the
relation $F(t)\propto{\dot{\varphi}}^{2}$ ceases to be valid. On
the other hand, the limit $V(\varphi) \, \gg \, \varrho_{pbh}$
leads to very simple solutions since this equation is separable
(note that $\varrho_{pbh}(t)= M n_{pbh}$) and consists of growing
solutions, holding for a finite time while $V(\varphi) \, > \,
\varrho_{pbh}$.

\bigskip
\subsection{Evaluating the fractional BH-mass variation for
quintessence flux.}
\bigskip

Let us evaluate the fractional change in the black hole mass (born
with an initial mass $M_{i}$) when the quintessence field alone
fuels the growth.

Writing $\Omega_{\varphi}={\varrho_{\varphi}\over{\varrho_{c}}}$
and noting that the contribution of the kinetic term is
${{\dot{\varphi}}^{2}\over{2}}$, then quite generally we have
$\Omega_{\varphi}={{1\over{2}}{\dot{\varphi}}^{2} +
V(\varphi)\over{\varrho_{c}}}$.

Therefore,
$\varrho_{c}\Omega_{\varphi}\sim{{\dot{\varphi}}^{2}\over{2}}$,
whenever ${1\over{2}}\dot{\varphi}^{2}>V(\varphi)$. When the
kinetic term dominates, we have
${dM\over{dt}}={27\pi\over{M_{pl}^{4}}}{M}^{2}\varrho_{c}(t)\Omega_{\varphi}(t)$,
where $t=t(z)$. This equation is easy to solve. Integrating it
from $t(z_{i})$ until $t(z=0)$, and from $M_{i}$ until $M(t)$
yields

\begin{equation}
\biggl({\Delta{M}\over{M}}\biggr)\sim{27\pi{M_{i}}\over{M_{pl}^{4}}}\int_{t(z_{i})}^{t(0)}
d{t}^{\prime}\varrho_{c}(t^{\prime})\Omega_{\varphi}(t^{\prime})
\,\,\, .
\end{equation}

Since the variation of the $\Omega_{\varphi}$ is expected to be
relatively small at low redshifts, and the quintessence term is
known to amount $\sim {0.7}$ today, we can always estimate the
kinetic contribution of the quintessence as a function of the
matter and radiation contributions. Neglecting as before the
curvature term we may rewrite the eq.(14) above as

\begin{equation}
\biggl({\Delta{M}\over{M}}\biggr)\sim{2.7\pi{M_{i}}\over{M_{pl}^{4}}}
\int_{z_{i}}^{0}dz^{\prime}\varrho_{c}(z^{\prime})
{dt^{\prime}\over{dz^{\prime}}} \,\,\, ;
\end{equation}

which, after substituting the relations between $z$ and $t$, in
terms of $H_{0}$, becomes

\begin{equation}
\biggl({\Delta{M}\over{M}}\biggr)(z_{i})\sim{M_{i}}H_{0}z_{i}
\,\,\, ;
\end{equation}

which is linear in $z_{i}$, as expected. The formula above is a
lower bound estimate, because we know that $\Omega_{\varphi}$
actually grows in that period. Then, it constitutes a minimal
bound, when that variation of the latter is small.

If we further parametrize the flux as
$F_{\varphi}={\varphi_{0}}{(t/t_{0})}^{n}$ for a generic index
$n$, we can obtain from the condition $\Omega_{0} < 1$

\begin{equation}
\varphi_{0} < \sqrt{\varrho_{c}(t_{0})}\sim{10}^{-23}{GeV}^{2}
\,\,\, .
\end{equation}

Supposing that the kinetic term dominates over the potential one,
and inserting this upper bound back into eq.(16) we obtain

\begin{equation}
\biggl({d\mu\over{d\tau}}\biggr)\leq{10}^{-22}{\mu}^{2}{\tau}^{n}
\,\,\, ;
\end{equation}

where $\mu \, = \, {M\over{M_{\odot}}}$ and $\tau \, = \,
t/t_{0}$.

In these particular parametrized power-law models, $n$ can be
positive, negative or null, and the results will differ for each
case. Consider $n > 0$ first. We can rise the following question:
when will the net mass variation be relevant compared to the
initial mass?. Let us suppose that we want to obtain a fractional
growth of $\sim{10}^{8}$. Inverting the above relation, the
cosmological time when this huge final mass will be achieved is

\begin{equation}
t(\mu_{i})\sim{10}^{30/(n+1)}{\mu_{i}}^{-1/(n+1)}t_{0} \,\,\, ;
\end{equation}

in other words, if the conditions $\Omega_{\varphi} < 1$ hold for
future times and $n$ is constant and positive, only in the very
remote future the scalar field accretion will be able to accrete
${10}^{8}$ times its initial mass. This future epoch of the
universe will be coasting (provided $\Omega_{total}=1$), and
dominated by big black holes that feed from the quintessential
energy in a very slow way. For $n < 0$, these conditions become
much more severe, and we do not get any substantial growth unless
very strong fine tuning had happened in the remote past. These
conclusions are not likely to change for realistic (or more
complete models) if we find the general background flux solutions
given by the Friedmann´s equations, since the fractional mass gain
is

\begin{equation}
\biggl({\Delta{\mu}\over{\mu}}\biggr)\sim{10}^{-22}{\mu_{i}}
\int_{\tau_{i}}^{\tau}{d\tau[f(\tau)]} \,\,\, ;
\end{equation}

and $f(\tau)$ is a consistent solution of the complete set of
equations. This is because it is very difficult to cancel the tiny
coefficient ${10}^{-22}$, which arises quite generally in the
consideration of the cosmological accretion. That is why we state
that extreme BH growth can arise only in very peculiar situations,
and why we believe that this is very difficult to achieve.

One can also invert the reasoning above and evaluate the density
of radiation/quintessence in remote past which would have produced
$({\Delta{M}\over{M}}) \gg 1$ and arrive to similar conclusions,
namely that even given the most favorable conditions the
quintessence would not be able to cause a huge growth of PBHs.

\section{Extreme fine-tuning for the growth of black holes
immersed in a quintessence background flux}

Using the critical mass (see \cite{CuHo1} for discussion and
references) as a tool let us show now that any flux of the form
$F_{\varphi} \propto {t}^{-n}$, with $n
> 1$ does not lead to a substantial growth for any black hole,
unless we require a strong fine-tuning between the formation-time
and the initial mass of the black hole (with all other parameters
kept fixed). This functional parametrization is specially suitable
for most applications and permits to make a quick contact with
standard cosmological evolutions.

The dynamical equation in the accretion regime is

\begin{equation}
\biggl({dM\over{dt}}\biggr)=CM^{2}M_{pl}^{4}{(t/t_{*})}^{-n}
\,\,\, ;
\end{equation}

where $C$, and $t_{*}$ are constants. Plain integration yields

\begin{equation}
\int_{M_{i}}^{M(\eta)}{dM\over{M^{2}}}={
C{M_{pl}}^{4}\over{E_{*}}}\int_{\eta_{i}}^{\eta}d\eta{\eta}^{-n}
\,\,\, ;
\end{equation}

with $\eta = (t/t_{*})$ and $E_{*}={t_{*}}^{-1}$.

The l.h.s. is just

\begin{equation}
{1\over{M_{i}}}-{1\over{M(\eta)}} \,\,\, ;
\end{equation}

and the r.h.s. is

\begin{equation}
{C{M_{pl}}^{4}\over{(n-1)E_{*}}}
\biggl[{{1\over{\eta_{i}^{n-1}}}-{1\over{\eta^{n-1}}}}\biggr]
\,\,\, .
\end{equation}

Then, the solution $M(\eta)$ is

\begin{equation}
{1\over{M(\eta)}}=
{1\over{M_{i}}}\biggl[1-{CM_{i}{M_{pl}}^{4}\over{(n-1)E_{*}}}
\biggl({{1\over{\eta_{i}^{n-1}}}-{1\over{\eta^{n-1}}}}\biggr)
\biggr] \,\,\, .
\end{equation}

Considering $n > 1$ we obtain for asymptotic times, i.e $\eta \gg
\eta_{i}$

\begin{equation}
M(\eta)={M_{i}\over{\biggl[1-{C{M_{pl}}^{4}\over{(n-1)E_{*}}}
{{M_{i}\over{\eta_{i}^{n-1}}} \biggr]}}} \,\,\, .
\end{equation}

It is clear that the denominator is much smaller than $1$ (and
positive) at the expense of a very strong fine tuning between
$M_{i}$ and $t_{i}$. Let us evaluate a definite numerical example.
Suppose that the growth factor must be of order ${10}^{8}$ as
before, i.e the denominator

\begin{equation}
1-{C{M_{pl}}^{4}\over{(n-1)E_{*}}}
{{M_{i}\over{\eta_{i}^{n-1}}}}={10}^{-8} \,\,\, .
\end{equation}

Then, the formation time $t_{i}$ must have been very fine-tuned
into the function above (in one part by several millions) in order
to get a huge growth factor. If this did not happen then either
the growth factor falls off quickly to $1$ or the solution is
unphysical. Note that this strong conclusion is very general, and
it is independent of the origin of the flux through the black hole
horizon, i.e. the result is valid for the radiation-dominated era,
in which the flux was $c\varrho_{rad}(t)\propto{t}^{-2}$ or a
quintessence-dominated era when
${\dot{\varphi}}^{2}\propto{t}^{-2}$ (see Ref.\cite{CuHo1}).

\section{Further constraints on PBH growth from the Generalized
Second Law and the Holographic Bound}

The irreversible process of quintessence disappearance into a
black hole is just a particular example of the Generalized Second
Law of thermodynamics (hereafter GSL) formulated in the '70s by
Bekenstein \cite{Jer} and other following works
\cite{Bigatti,Kalo}.

The GSL is the statement that the total entropy (black holes plus
any other fields) never decreases with time

\begin{equation}
\Delta{S}_{total} = \Delta{S_{bh}} + \Delta{S_{matter}} +
\Delta{S_{\varphi}} + \Delta{S_{radiation}} > 0 \,\,\, ;
\end{equation}

where

\begin{equation}
S_{bh}(N,M)\sim{10}^{77}N{(M/M_{\odot})}^{2} \,\,\, ;
\end{equation}

counts for the entropy of $N$-black holes, $S_{matter}$ and
$S_{radiation}$ is the normal matter and radiation entropies, and
$S_{\varphi}$ is the entropy of the scalar field. A step forward
towards the use of entropy as a tool in cosmology is the so-called
Holographic Principle (HP). The Holographic Principle
\cite{Bigatti,Kalo,Bou}, is the statement that all
entropy/information contained in the physical universe can be
attributed to its boundary. In numerical terms, the upper bound to
total entropy (known as the Holographic Bound) is

\begin{equation}
S_{max}(t)<{A(t)\over{4L_{pl}^{2}}} \,\,\, ;
\end{equation}

where $A(t)$ is the boundary area of the physical system at $t$.
If we assume that the HP can be applied to the cosmological
models, then it is natural to choose the cosmological particle
horizon to bound the area and write the maximum entropy as

\begin{equation}
S_{hp}(t)\sim{8\times{10}^{121}}{(t/t_{0})}^{2} \,\,\, .
\end{equation}

If the inequality $S_{bh}(N,M) < S_{hp}(t)$ has to be satisfied,
the accretion of quintessence should adjust in such a way that a
black hole with initial mass $M_{i}$ may grow to values
$M_{f}>{10}^{6}M_{\odot}$, without violating the Holographic
Bound. Therefore, an upper bound to the local flux due to this
global requirement is obtained.

It is clear that this energy input strongly depends on the initial
mass of the PBH $M_{i}$ and needs to be very large if the black
hole was initially small. In addition, if the accretion is very
high, the black hole that was initially below the Holographic
Bound will blow that bound at some point. Then, to keep these
black holes below the Holographic Bound at any time, the
constraint

\begin{equation}
\dot{S}_{smbh}(t) < \dot{S}_{hp}(t) \,\,\, ;
\end{equation}

must be satisfied, together with $S_{smbh}(t_{i})<S_{hp}(t_{i})$
at the formation time $t_{i}$. When both conditions are required
to hold, an absolute upper bound to the flux onto the $N$
accreting black holes can be established. It comes out to be
$F_{\varphi} <
{{M_{\odot}^{2}}\over{t_{0}^{2}}}
\biggl[{10^{44}t\over{CNM^{3}(t)}}\biggr]$,
or numerically

\begin{equation}
F_{\varphi}(t)<{4.8\times{10}^{-5}}
{(t/s){h_{0}}^{2}\over{(N/{10}^{11}){(M/M_{*})}^{3}}}g{cm}^{-2}{s}^{-1}
\,\,\, ;
\end{equation}

with $M_{*}\sim{10}^{8}M_{\odot}$.

If the total number $N$ of black holes increases, the flux
$F_{\varphi}(t)$ would be correspondingly smaller, in order to
keep all black holes growing at moderate rate and the total
entropy smaller than $A_{hp}/4$.

A general and rigorous proof that the Holographic Bound can be
applied to the cosmological models is lacking now, but if we
proceed with these ideas we can get some insight about the local
and global properties of black holes in realistic models of the
universe, although the full meaning of eq.(44) is obscure yet.
Some problems of these methods in cosmology are discussed in
Refs.\cite{Kalo} and \cite{Bou}. Upper bounds to black hole
abundances have been obtained using these ideas \cite{CuHo2}.

Quite regardless of the use of the Holographic Bound, we expect
that when the mass-energy of black holes is comparable to term
$V(\varphi)$ in the Friedmann equation, the system response would
be to deplete the quintessence flux onto the black hole horizon by
decreasing ${\dot\varphi}^{2}$. A complete study of these features
will be published elsewhere \cite{CuHo3}.

\bigskip
\section{Conclusions}
\bigskip
We have shown in this work that, generally speaking, the
quintessence flux ${\dot{\varphi}}^{2}$ must decrease slower than
${t}^{-2}$ for PBHs to grow at all, and to stay constant or
increase in time for substantial accretion to occur. This may be
relevant for some PBHs achieving the SMBH condition in a short
time, but only at the expense of fine-tuning in the parameters.
Several options for the quintessence potential $V(\varphi(t))$
have been proposed in the literature, some of which can lead to
high flux to feed PBHs (i.e. leading to large positive $\dot{M}$)
for long times. One analytical solution of this class leading to
${\dot{\varphi}}^{2} = constant$ has been worked out as an
example.

If the flux onto the black holes is constant, all them initially
above the critical mass will grow until the term $\varrho_{PBH}$
approaches the energy-density in the cosmic background. From this
point, these black holes would not keep growing at high rates, and
the more complicated calculations have not been performed for this
regime.

In Section 3 we showed that if the flux behaves as
$F(t)\propto{t}^{-2}$, we can prove that all solutions with
asymptotic growth ($M(t) \gg M_{i}$) depend of an initial
fine-tuning between the initial masses $M_{i}$ and their formation
at $t=t_{i}$. Other fluxes that decrease more slowly (or even
increase) than ${t}^{-2}$ may lead to very big black holes but
only at $t \gg t_{i}$, a condition that is not favorable for
quintessence accretion which is unimportant at high redshifts and
thus does not have enough time to happen.

In Section 4 we have invoked Holographic Bound arguments to show
that a bound on the quintessence flux can be obtained quite
generally, although these arguments would only hold if a rigorous
proof of the existence of a bound is obtained. In spite of the
theoretical efforts, this has not been achieved and therefore the
limits must be considered as speculative.

It should be kept in mind that models others than the one
discussed here can be constructed to produce a population of SMBHs
starting from seed PBHs. For example, accretion in a brane-world
high-energy phase has been recently studied by Guedens, Clancy and
Liddle \cite{GCL} and Majumdar \cite{Mad}, which show that a
substantial growth in which $\dot{M} \propto \, M/t$ is allowed.
It may be possible to arrive to the end of the high-energy phase
with very massive black holes, although the full consequences of
this type of scenarios are yet to be explored.

\section{Acknowledgments}

The authors wish to thank the S\~ao Paulo State Agency FAPESP for
financial support through grants and fellowships. J.E.H. has been
partially supported by CNPq (Brazil). We also acknowledge
scientific advise from Prof. Elcio Abdalla.

\section{Appendix: the critical mass in radiation+quintessence environments}

A black hole surrounded by radiation/quintessence would be either
absorbing energy or evaporating, according to the sign of the
difference $T_{pbh} - T_{environment}$, just like any other
thermodynamical system. Consideration of the complete differential
equation describing the black hole mass evolution

\begin{equation}
\biggl({dM\over{dt}}\biggr)=-{A\over{M^{2}}}+C{M}^{2}F_{total}(t)
\,\,\, ;
\end{equation}

is needed (where $F_{total}(t)$ represents the total flux of
energy through the black hole horizon (that is, the addition of
fluxes for any form of energy that falls through the black hole)
at cosmic time $t$). For a given $F_{total}$ the condition
${dM\over{dt}} = 0$ defines a parameter (the {\it critical mass}
$M_{c}(t)$) separating the absorbing and evaporating regimes. The
most interesting cases arise when the flux is dominated by
radiation or quintessence. We shall discuss the cases separately
to highlight some interesting differences.

\subsection{Radiation-dominated environment}

The case of pure radiation $F_{total} \equiv
F_{rad}(t)=c\varrho_{rad}(t)$ has been analyzed in \cite{CuHo}. By
solving the algebraic equation ${dM\over{dt}} = 0$, the critical
mass is found to be

\begin{equation}
M_{c}(T_{rad})={\theta\over{T_{rad}(t)}}
\sim{{10}^{26}g\over{(T_{rad}/T_{0})}} \,\,\, ;
\end{equation}

where $\theta=M_{pl}{(A/27{\pi}a_{*})}^{1/4}$,
$\varrho_{rad}(T)=a_{*}T^{4}$ and $T_{0}\sim{3K}$ is the present
temperature of the CMBR.

Then, if we ignore other possible sources of energy-matter to fuel
their growth, we conclude that any black hole with mass $ >
{10}^{26} g$ is today in the absorption regime. The critical mass
has evolved with time, it was $M_{c}\sim{10}^{16}g$ when the
universe was $t \sim {1s}$ old. Therefore, any PBH with initial
mass greater than the latter value formed within the
radiation-dominated era, was initially above the equilibrium
condition, and the Hawking evaporation was irrelevant for it.
Apart from fine details of the absorption process, the cross
section is expected to be valid quite generally (see Refs.
\cite{CuHo} and \cite{CuHo1} for a full discussion).

\subsection{Quintessence-dominated environment}

If we consider the quintessential field as the dominating fuel,
the balance equation for the black hole mass ${dM\over{dt}} = 0$
becomes

\begin{equation}
\biggl({dM\over{dt}}\biggr)_{total}=
-{A\over{M^{2}}}+{C{M}^{2}\over{2}}{{\dot{\varphi}}^{2}} \,\,\, .
\end{equation}

In order to simplify the expressions, let us define the
dimensionless ratio $A={\sigma}M_{pl}^{4} \, \sim \,
{1.1\times}{10}^{-3}$. Thus, the critical mass for this case is
expressed as

\begin{equation}
M_{c}(t)={0.06{M_{pl}}\over{{\Xi (t)}^{1/4}}}
\sim{5.6\times{10}^{24}g}{\biggl[{{\dot{\varphi}}^{2}
\over{2\varrho_{c}(t_{0})}}\biggr]}^{-1/4} \,\,\, ;
\end{equation}

where we have defined the dimensionless function $\Xi
(t)={{\dot{\varphi}}^{2}\over{2{M_{pl}}^{4}}}$.

The expression shows that if the quintessential density is of
order of the critical density today, the corresponding critical
mass is about ${10}^{25}g$. This is not very different than the
value corresponding to pure radiation dominance, a fact directly
related to the assumption $\Omega_{\varphi} \, \sim \,
\Omega_{crit}$.

Let us show how the constant flux case described in Section 2
looks like in terms of the critical mass. The critical mass is a
constant given by

\begin{equation}
M_{c}(t)\sim 5.6\times{10}^{24} \, g
{{\biggl[{B\over{\varrho_{c}(t_{0})}}\biggr]}}^{-1/4} \,\,\, .
\end{equation}

Therefore, only initial masses greater than this value can be
described by growing models with constant quintessence through the
horizon. Fig. 2 displays the behavior of black holes in this
situation according to their initial mass, see caption for
details. More general models can be studied. For instance, let us
briefly discuss the case of a power-law flux
$F(t)=M_{pl}^{4}{(t/t_{*})}^{n}$ with $t_{*}$ a reference time. In
the case $n < 0$, we obtain for the critical mass

\begin{equation}
M_{c}(t)\sim{(\sigma/27\pi)}^{1/4}{M_{pl}}{(t/t_{*})}^{|n|/4}
\,\,\, ;
\end{equation}

and this critical mass {\it increases} with increasing time. Since
the ratio of the absorption term to the evaporation term is just

\begin{equation}
-{\dot{M}_{abs}\over{\dot{M}_{evap}}}={(M/M_{c})}^{4} \,\,\, ;
\end{equation}

(valid for any form of the accreting flux) then, if we follow the
evolution of a PBH mass whose initial mass $M_{i}$ was slightly
larger than $M_{c}(t=t_{i})$, then we can ignore the Hawking term
on its evolution until the PBH mass crosses the critical mass
curve. The error introduced to obtain the crossing time $t_{c}$ is
small. Thus we can describe the PBH mass evolution as

\begin{equation}
{1\over{M_{i}}}-{1\over{M_{c}(t_{c})}}={27\pi{t_{*}}\over{(n-1)}}
\biggl[{1\over{\eta_{i}^{n-1}}}-{1\over{\eta_{c}^{n-1}}}\biggr]
\,\,\, ;
\end{equation}

where $\eta_{i}=(t_{i}/t_{*})$ and $\eta_{c}=(t_{c}/t_{*})$. Since
$t_{c} \gg t_{i}$ in general, the second term is always
negligible.

\bigskip
EDITOR: PLEASE PUT FIG.1 HERE!!!!!!!!
\bigskip

Solving eq.(22)for the crossing time yields

\begin{equation}
t_{c}(M_{i},t_{i})\sim{t_{*}}{\biggl({M_{i}\over{M_{pl}}}\biggr)}^{4/n}
{\biggl({\sigma\over{27\pi}}\biggr)}^{-1/n}
{\biggl[{1-{27\pi{M_{i}}t_{*}\over{n-1}}{(t_{i}/t_{*})}^{1-n}}\biggr]}^{-4/n}
\,\,\, .
\end{equation}

Then, in a finite time, this PBH will achieve an instantaneous
equilibrium at the critical mass crossing, and immediately proceed
into the evaporating region.

In the case $n > 0$, the critical mass is

\begin{equation}
M_{c}(t)\sim{M_{pl}}{(\sigma/27\pi)}^{1/4}{(t/t_{*})}^{-|n|/4}
\,\,\, .
\end{equation}

And this function {\it decreases} with increasing time. Suppose
that a PBH has been formed with initial mass smaller than the
critical mass at $t_{i}$. Then, for a long time this PBH will loss
energy until its complete evaporation (if we ignore other quantum
corrections likely to arise near the Planck scale). However, if
its mass reached the critical mass at value $M_{c}(t_{c}) >
M_{pl}$, then the black hole achieved an instantaneous equilibrium
at $t_{c}$ (see below) and it will proceed into the {\it
absorbing} region, from which it will be growing increasingly
faster (Fig. 3). We note that at $M=M_{c}$ we have
$\ddot{M}(M=M_{c},t_{c})\propto{M^{2}{(t/t_{*})}^{n-1}}>0$, and it
will be increasing for $n > 1$. Now, if a PBH had already formed
below the critical mass at $t_{i}$, and is not much smaller than
it we can solve the mass differential equation to find the
crossing time

\begin{equation}
t_{c}(M_{i})\sim{t_{*}{\biggl({\sigma\over{27\pi}}\biggr)}^{1/n}
{\biggl({M_{pl}\over{M_{i}}}\biggr)}^{4/n}} \,\,\, .
\end{equation}

This solution is numerically valid only when $t_{c}(M_{i}) \ll
t_{evap}\sim{t_{0} {({M_{i}/M_{haw}})}^{3}}$.

From $t_{c}(M_{i})$, this black hole will be accreting and its
mass will be growing. These considerations illustrate the role of
the critical mass for the evaluation of the fate of PBHs. Any
statement about growth/evaporation is necessarily time-dependent
and must be carefully done to preclude a misunderstanding of the
physical situation.

\bigskip
EDITOR: PLEASE PUT FIG.2 HERE!!!!!!!!
\bigskip

In a completely generic case $F_{total}=F(t)$ we obtain

\begin{equation}
\ddot{M}(M,t)=\biggl[{{2A\over{M^{3}}}+2CBM}\biggr]\dot{M}(M,t)+CM^{2}(t)\dot{F}(t)
\,\,\, ;
\end{equation}

and the fate of these black holes will depend on the sign of
$\dot{F}(t)$ as expected. In general, $\ddot{M}(M=M_{c},t)$ is not
zero, and a similar conclusion holds for higher derivatives as
well. Therefore, we conclude that very special conditions have to
be fulfilled for black holes to evolve in equilibrium with the
environment, that is, without losing or gaining mass.

\subsection{Quintessence/radiation environment}

If both radiation and quintessence are considered, then the
critical mass becomes

\begin{equation}
M_{c}(t)\sim{{{10}^{26}g\over{(T_{rad}/T_{0})}}}
{\biggl[{1+{{\dot{\varphi}}^{2}\over{\varrho_{rad}}}}}\biggr]^{-1/4}
\,\,\, ;
\end{equation}

to be solved together with eqs.(4-6). In general, the behavior of
$\varrho_{rad}$ would not be $\propto \, t^{-2}$, which happens
only when the radiation dominates all forms of energy. In this
case, the first derivative of $\dot{M}(M,t)$ at $M=M_{c}$ is given
by

\begin{equation}
\ddot{M}(M=M_{c},t)={C{M}^{2}}\bigl({4\dot{T}T^{3}
+\dot{\varphi}\ddot{\varphi}}\bigr) \,\,\, ;
\end{equation}

and this term is in general not zero unless
$\dot{\varphi}\ddot{\varphi}=-4{\dot{T}}T^{3}$. Since radiation
and quintessence would have to conspire for that, equilibrium of
PBHs and the environment would not be possible. In scenarios where
$V(\varphi)>>\varrho_{rad}$ PBHs can stay in equilibrium, actually
one of these is the constant flux model already discussed
${\dot{\varphi}}^{2}=2B=cte$.

Now, we can evaluate the redshift dependence in these terms. In
any way, $\varrho_{rad}(z)\propto{(1+z)}^{4}$, regardless of
whether the quintessence dominates. We can cast the eq.(47) above
as

\begin{equation}
M_{c}(z)\sim{10^{26}g \over{{\biggl[{(1+z)}^{4}
+{1\over{\varrho_{rad}(0)}}{\biggl({{d\varphi}\over{dz}}\biggr)}^{2}{\dot{z}^{2}}}\biggr]}
^{1/4}} \,\,\, .
\end{equation}

It is easy to conclude that if ${1\over{{(1+z)}^{4}}}
{1\over{\varrho_{rad}(0)}}{\biggl({{d\varphi}\over{dz}}\biggr)}^{2}{\dot{z}^{2}}
\propto{(1+z)}^{m-4}$ the following behaviors are possible. If $m
> 4$ the quintessence will determine the critical mass even at high
redshifts. For $m<4$, the opposite case happens, and there will be
a redshift $z_{l}$ from which the radiation will determine the
critical mass. It will be true at
$(1+z_{l})\sim{\biggl[{({d\varphi\over{dz}})}^{2}{\dot{z}}^{2}\biggr]}^{1/4}$,
evaluated at $z_{l}$ also. It depends (as it should be) on the
time derivative of the quintessential field, and the complete
solution must be solved when we choose some form of $V(\varphi)$.
It is important note that when the term $V(\varphi)$ is relevant,
the accretion into black holes can be null in some cases even when
$\Omega_{\varphi}$ is big. We can write
$\Omega_{\varphi}={{1\over{2}}{\dot{\varphi}}^{2}+V(\varphi)\over{\varrho_{c}}}$.
Then, when the first term is important, the quintessence dominates
and the black hole accretion (from it) is important also, but the
inverse is not true. We can choose other models with $V(\varphi)$
leading to ${1\over{2}}{\dot{\varphi}}^{2}=0$, and the
quintessence dominates without some secondary effect onto black
holes with any mass. The third point now is. When the form of
$V(\varphi)$ leads to a non-null kinetic term, the magnitude of
this term depends also on the magnitude of $V(\varphi)$ (obtained
by solving the dynamics) and the accretion onto black holes will
depends on the parameters that enter into $V(\varphi)$-
definition. Finally, the fourth point is that the z-dependence in
$\Omega_{\varphi}(z)={{1\over{2}}{\dot{\varphi}}^{2}(z)+V(\varphi(z))\over{\varrho_{c}(z)}}$
reflects in each term. Today we know that
$\Omega_{\varphi}(z\sim{0})\sim{0.7}$, but the attractor features
of the solutions indicate that when $z \geq 1$, the magnitude of
the terms above were not so much below the unity also, although
the exactly dependence and values are not known yet. In terms of
time, from $z=10$ to $z=0$ there was enough time to accretion into
black holes, if the others conditions were satisfied together
also. Moreover, there are indications that the quintessence varies
with $z$ in a very slow way if at all, see \cite{SN}.

\bigskip
\subsection{The behavior of $\ddot{M}(M=M_{c},t)$}
\bigskip

An interesting related question is whether PBHs formed with
$M_{i}=M_{c}$ can stay in equilibrium without evaporating or gain.
We give a brief assessment of the general features of the
solutions $M(t)$. Given the general form of $\dot{M}(M,t)$ we may
express $\ddot{M}(M_{c},t)$ in the following way

\begin{equation}
\ddot{M}(M_{c},t)=CM^{2}(t)\dot{F}(t) \,\,\, ;
\end{equation}

in terms of the mass and the behavior of the flux at its horizon.
Since $M = M_{c}$, then it will be simplified to
$CM_{c}^{2}(t)\dot{F}(t)$ and it is enough know the evolution of
the critical mass and flux. The true behavior of each black hole
mass is very complicated even in the simply cases above, but
through the critical mass and its derivatives, we may infer the
black hole mass at asymptotic times. Then, the algebraic sign and
derivatives of $\ddot{M}(M_{c},t)$ will determine the behavior of
the solutions.

Applying these definitions to the constant flux case we obtain
${\dot{M}}(M=M_{c},t)=0$ on the critical mass. The first
derivative of ${\dot{M}}(M,t)$ is just
${d\over{dt}}{\biggl(-{A\over{M^{2}}}+{C{M}^{2}B}\biggr)}$,
rewritten as

\begin{equation}
\ddot{M}(M,t)=\biggl[{{2A\over{M^{3}}}+2CBM}\biggr]\dot{M}(M,t)
\,\,\, .
\end{equation}

Since $\ddot{M}(M=M_{c},t)=0$, equilibrium is possible on the
critical mass and the black holes will retain their initial
masses. This result is valid as long as we ignore the radiation
compared to the quintessence field. Note that all the higher
derivatives of $\dot{M}(M,t)$ are null at $M_{c}$ only if the flux
through the horizon is constant.

For the example mentioned in Section 3 of a power-law flux
$F(t)=M_{pl}^{4}{(t/t_{*})}^{n}$. The critical mass is
$M_{c}(t)={\alpha\over{F^{1/4}(t)}}$ with $\alpha={(A/C)}^{1/4}$
and the second derivative $\ddot{M}$ reads

\begin{equation}
\ddot{M}(M_{c},t)={nC{\alpha}^{2}M_{pl}^{2}\over{t_{*}^{n}}}{(t/t_{*})}^{n/2-1}
\,\,\, .
\end{equation}

From eq.(31) above, it is clear that we have a branching of the
behavior for the solutions. The first branch corresponds to $n >
2$ and in the limit $t\rightarrow{\infty}$ we obtain
$\ddot{M}(M_{c},t\rightarrow{\infty})=\infty$. On the other hand,
$n < 2$ yields $\ddot{M}(M_{c},t\rightarrow{\infty})= 0$ These
cases are physically very different, because the function
$\dot{M}(M,t)$ is related to the energy that crosses the horizon
event, and its derivative is sensitive to changes in the
asymptotic limit. The cases $0 < n < 2$ correspond to solutions
showing moderate growth and the case $n > 2$ to very fast growth
of the black holes. The cases $n < 0$ are the ones for which the
critical mass grows, the behavior of black holes in this latter
case is determined by the initial condition $M_{i}$.

\vfill\eject

\noindent
Figure captions.

\begin{figure}
Fig.1. The critical mass for a constant quintessence flux. The
critical mass is represented by the horizontal line labelled in
the graph. A black hole may either grow or evaporate according to
the value of its initial mass $M_{i}$ above or below the critical
mass respectively. Instantaneous equilibrium of a black hole not
growing nor evaporating is possible if its mass is precisely the
critical value, as discussed in the text. Note that growing is
expected to saturate when the primordial black hole density
approaches the quintessence field density.
\end{figure}

\begin{figure}
Fig. 2. The same as in Fig. 1 for a power-law quintessence flux of
the form  $F(t)=M_{pl}^{4}{(t/t_{*})}^{n}$, with $n > 0$. The
black holes suffer a delay of their evaporation and may start to
grow if they manage to cross the critical mass curve which
decreases with increasing time. This case illustrates how
different acceptable models of quintessence have to be analyzed
individually to address the fate of PBHs possibly formed at early
times .
\end{figure}

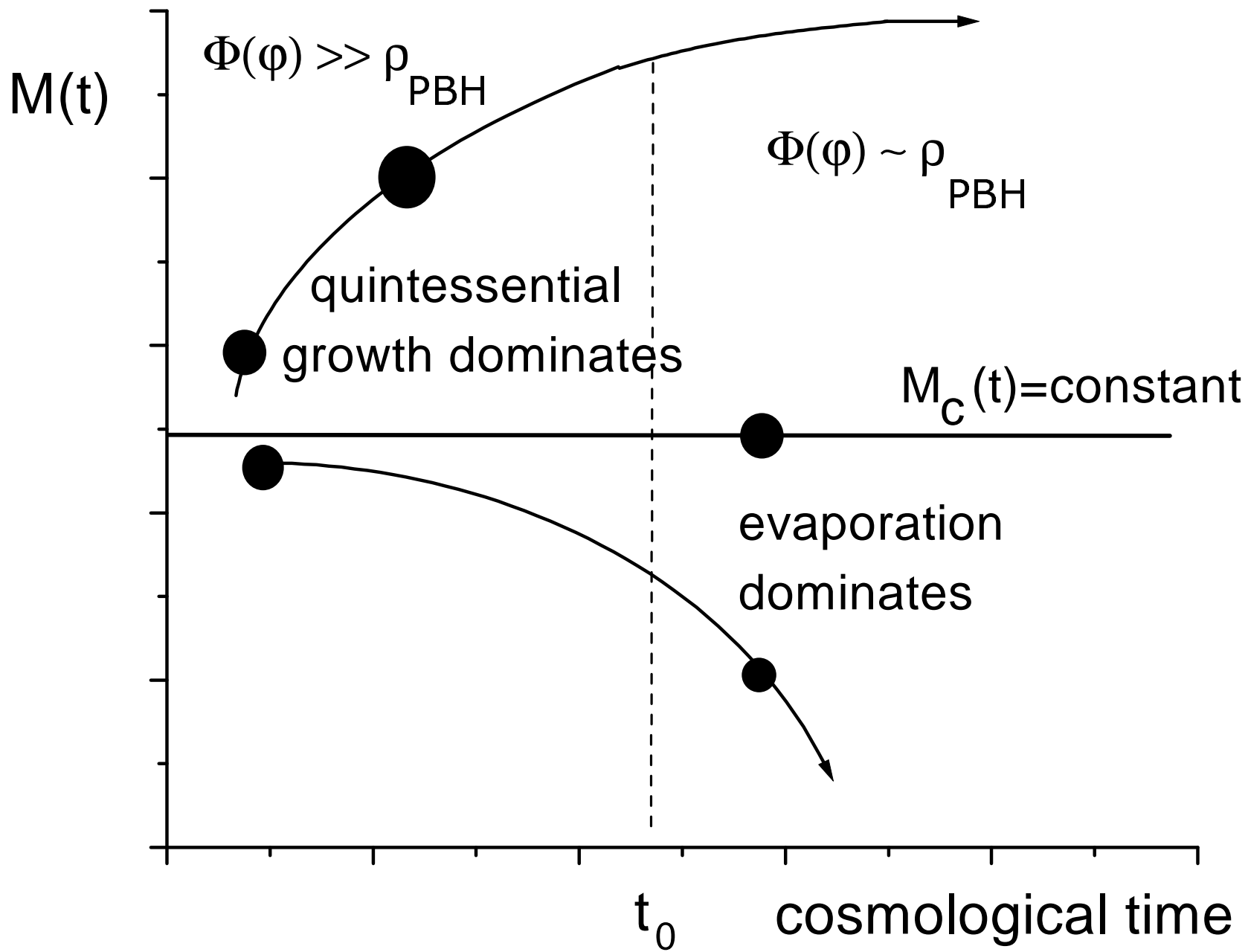
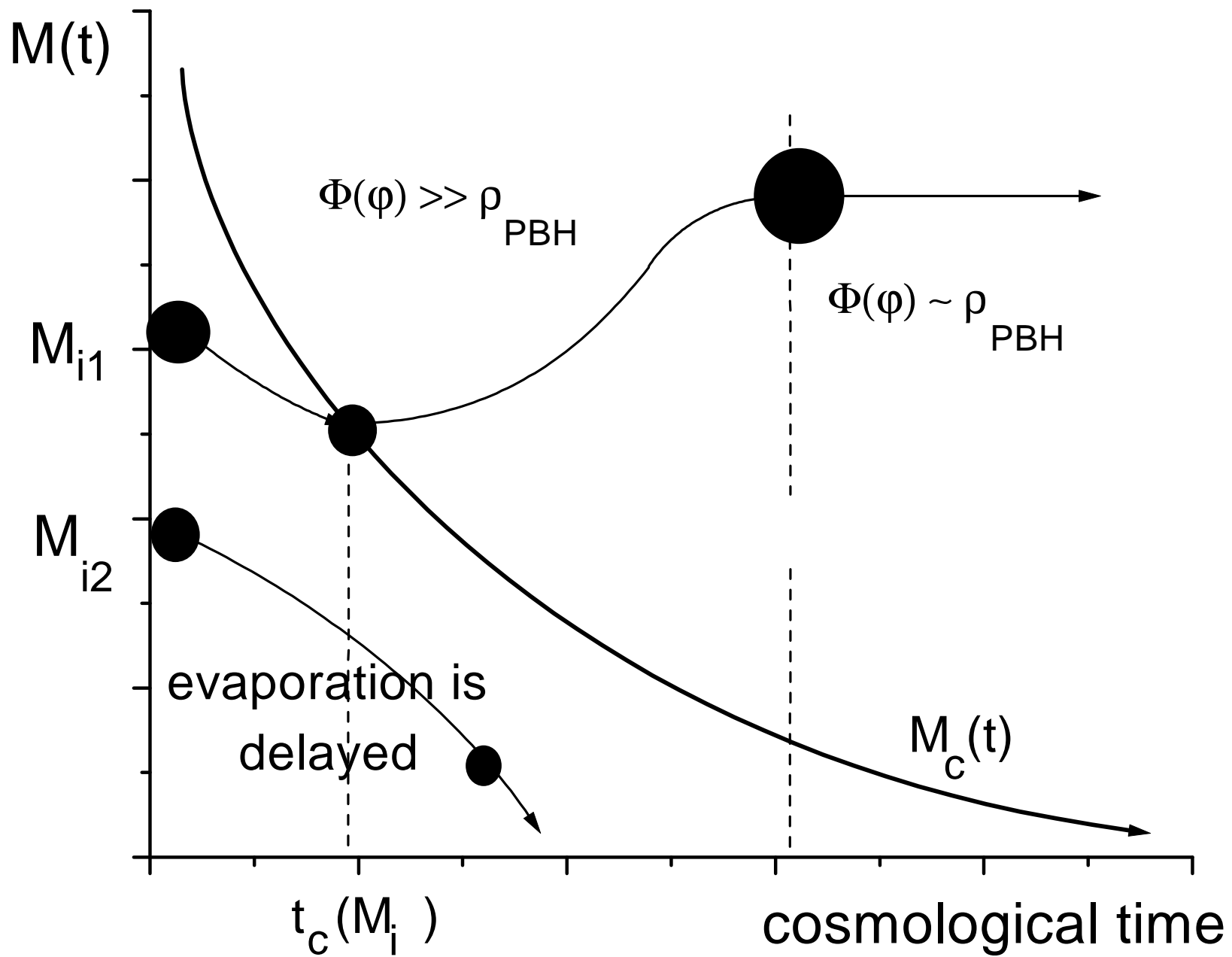
\end{document}